\documentclass[twocolumn,english,aps,prd,reprint,floatfix,notitlepage,footinbib,preprintnumbers,superscriptaddress,altaffilletter]{revtex4-1}
\pdfoutput=1
\usepackage{lmodern}

\usepackage[T1]{fontenc}
\usepackage[latin9]{inputenc}
\usepackage{geometry}
\usepackage{subfigure,lmodern, amsmath,amssymb, graphicx, pifont, adjustbox, bm, xcolor}
\usepackage{amsfonts}
\usepackage{comment}
\usepackage{mathtools}
\usepackage{float}
\usepackage{slashed}
\usepackage{ragged2e}
\usepackage{array}
\usepackage{hhline}

\makeatletter\g@addto@macro\bfseries{\boldmath}\makeatother

\usepackage{stackengine}
\usepackage{esint}
\usepackage[unicode=true,pdfusetitle,
 bookmarks=true,bookmarksnumbered=false,bookmarksopen=false,
 breaklinks=false,pdfborder={0 0 1},backref=false,colorlinks=true]
 {hyperref}
\hypersetup{
 pdfauthor={Nima Arkani-Hamed, Clifford Cheung, Carolina Figueiredo, and Grant N. Remmen},
 citecolor=black,linkcolor=black,urlcolor=black}

\newcommand{\appendixref}[1]{\hyperref[#1]{appendix~\ref{#1}}}
\def\equationautorefname~#1\null{eq.\,(#1)\null}
\usepackage{breakurl}
\usepackage{breakurl}
\usepackage[hang,flushmargin]{footmisc} 
\allowdisplaybreaks
\makeatletter

\usepackage{etoolbox}
\apptocmd{\thebibliography}{\justifying\setlength{\leftskip}{7.4mm}}{}{} 
 
 \usepackage{relsize}
\usepackage{babel}

\makeatletter
\def\simgt{\mathrel{\lower2.5pt\vbox{\lineskip=0pt\baselineskip=0pt
           \hbox{$>$}\hbox{$\sim$}}}}
\def\simlt{\mathrel{\lower2.5pt\vbox{\lineskip=0pt\baselineskip=0pt
           \hbox{$<$}\hbox{$\sim$}}}}
\makeatother

\usepackage{changepage}

\newcommand{\be}{\begin{equation}}
\newcommand{\ee}{\end{equation}}
\newcommand{\bea}{\begin{eqnarray}}
\newcommand{\eea}{\end{eqnarray}}
\newcommand{\Fig}[1]{Fig.~\ref{#1}}

\newcommand{\Eq}[1]{Eq.~(\ref{#1})}
\newcommand{\Eqs}[2]{Eqs.~(\ref{#1}) and (\ref{#2})}

\newcommand{\eq}[2]{\be\begin{aligned}#1 \label{#2}\end{aligned}\ee}

\renewcommand{\ao}{\alpha_0}
\newcommand{\os}{\overstar\omega}
\newcommand{\Rs}{\overstar R}

\newcolumntype{P}[1]{>{\centering\arraybackslash}p{#1}}

\begin{document}

\preprint{CALT-TH 2023-051}

\title{Multiparticle Factorization and the Rigidity of String Theory}

\author{Nima Arkani-Hamed}
\affiliation{School of Natural Sciences, Institute for Advanced Study, Princeton, NJ 08540}
\author{Clifford Cheung}
\affiliation{Walter Burke Institute for Theoretical Physics, California Institute of Technology, Pasadena, CA 91125}
\author{Carolina Figueiredo}
\affiliation{Department of Physics, Princeton University, Princeton, NJ 08540}
\author{Grant N. Remmen}
\affiliation{Center for Cosmology and Particle Physics, Department of Physics, New York University, New York, NY 10003}
    
\begin{abstract}
\noindent  Is string theory uniquely determined by self-consistency? Causality and unitarity seemingly permit a multitude of putative deformations, at least at the level of two-to-two scattering.  Motivated by this question, we initiate a systematic exploration of the constraints on scattering from higher-point factorization, which imposes extraordinarily restrictive sum rules on the residues and spectra defined by a given amplitude.  These bounds handily exclude several proposed deformations of the string: the simplest ``bespoke'' amplitudes with tunable masses and a family of modified string integrands from ``binary geometry.'' While the string itself passes all tests, our formalism directly extracts the three-point amplitudes for the low-lying string modes without the aid of worldsheet vertex operators.

\end{abstract}
\maketitle

\preprint{}

\maketitle

\noindent {\bf Introduction.}---The scattering amplitudes of string theory exhibit a litany of seemingly impossible features simultaneously.  On the one hand, they offer a proof of principle for how to consistently extrapolate quantum gravity to ultrahigh energies, past the regime of validity of general relativity.   Miraculously, they achieve all this while preserving causality, which mandates that high-energy elastic scattering is polynomially bounded in the center-of-mass energy at fixed impact parameter~\cite{Martin:1965jj,Jin:1964zza,Eden:1971fm,Adams:2006sv,Haring:2022cyf,Arkani-Hamed:2020blm,Camanho:2014apa}.  Furthermore, string amplitudes make crucial use of higher-spin states, which are famously inconsistent in isolation  but appear together in an infinite tower that elaborately conspires to maintain unitarity. 

\begin{figure}[t]
	\centering
	\includegraphics[width=5.cm]{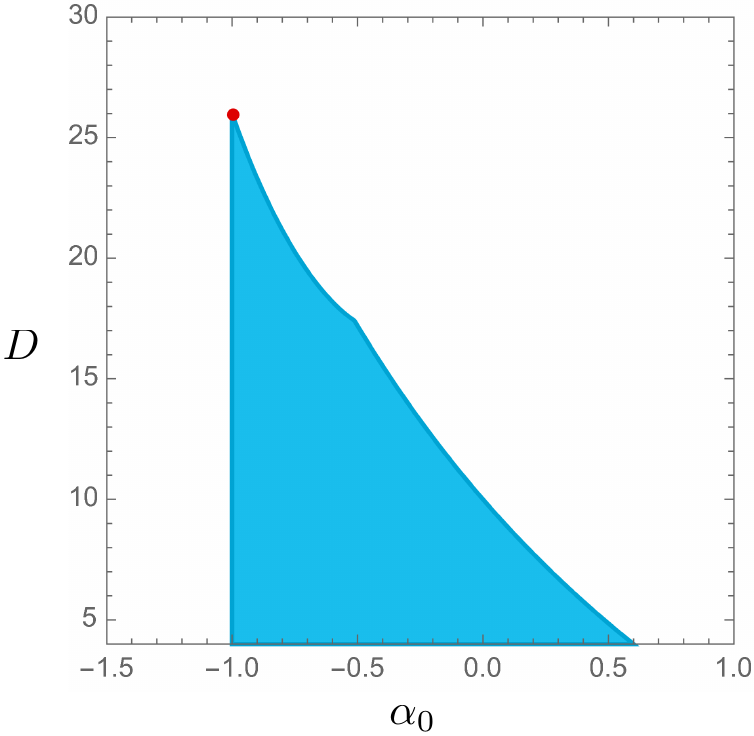}\vspace{-2mm}
	\caption{Partial wave unitarity constrains the four-point open string amplitude $\Gamma({-}s{+}\alpha_0)\Gamma({-}t{+}\alpha_0)/\Gamma({-}s{-}t{+}{2}\ao)$ with Regge intercept $\ao$ in $D$ spacetime dimensions to the allowed region in blue.  The bosonic string corresponds to the cusp at $\ao\,{=}\,{-}1$ and $D\,{=}\,26$.}
	\label{fig:string}
\end{figure}

These conditions are so stringent that one might wonder whether string amplitudes are {\it uniquely determined} by this panoply of constraints.   
Recent years have witnessed a surge of interest in this question in the very simplest context of four-point scattering~\cite{Camanho:2014apa,Caron-Huot:2016icg,Arkani-Hamed:2020blm,Arkani-Hamed:2022gsa,Alday:2022xwz,Berman:2023jys,Cheung:2022mkw,hypergeo}.  Even in just the past few months, a dizzying array of such proposals has made an appearance in the literature~\cite{hypergeo,Cheung:2023uwn,Cheung:2022mkw, Sasha, Huang:2022mdb,Maldacena:2022ckr, Geiser:2022exp,Geiser:2023qqq,Bhardwaj:2022lbz,Bhardwaj:2023eus,Jepsen:2023sia}, some revisiting the seminal work of Ref.~\cite{Coon}. 
 Rather surprisingly, the broad conclusion has been negative: it is possible to deform the four-point Veneziano amplitude~\cite{Veneziano:1968yb} of the open string in many ways that still conform to these physical constraints.   Even at four points, partial wave unitarity still allows for a range of consistent parameters~\footnote{Naive application of partial wave unitarity would suggest that positivity at $k=1$ requires $\ao\leq 1/3$. However, for $\ao>1/3$ the exchanged $k=1$ mode requires less energy than the center-of-mass energy to go on-shell.  Hence, this factorization channel is inaccessible and the positivity bound does not apply.}, though the famous $D=26$ of string theory still takes on special significance, as shown in \Fig{fig:string}.

It is natural to ask whether this immense freedom persists beyond four-point scattering. 
However, the  $n$-point frontier of scattering has remained largely uncharted territory until the past year or so. 

For example, a vast class of generalizations of $n$-point string amplitudes has been recently proposed~\cite{Cheung:2023uwn} that exhibit a ``bespoke,'' or completely tunable, arbitrary spectrum.  Like in string theory, these bespoke amplitudes describe an infinite tower of higher spins exhibiting channel duality.  
At four points, there is evidence that specific models, at least in certain regions of parameter space, are causal and unitary. Furthermore, at $n$ point these amplitudes are consistent with factorization on the poles of the lowest-lying exchanged state.

Another salient development is a new approach to $n$-point scattering that applies to string theory but also quantum field theories like $\phi^3$ theory and Yang-Mills theory at all orders in the 't~Hooft expansion.  In this framework, $n$-point amplitudes are described by a  simple integral over a set of auxiliary $u$ variables that are the mathematical cousins of the usual moduli that define the Koba-Nielsen representation of string amplitudes.  Conveniently, this formalism accommodates a natural deformation of the $n$-point string amplitude through the introduction of a multiplicative form factor $P_n(u)$ into the integrand.  This function can be straightforwardly engineered so as to ensure factorization on massless poles.

The upshot of these results is a bit of whiplash: we have gone from the naive intuition that $n$-point string scattering is unique to an embarrassment of riches in putative consistent deformations!  This is exciting and disturbing in equal parts.  But is this freedom genuine?

\begin{figure}[t]
	\centering
	\includegraphics[width=8.4cm]{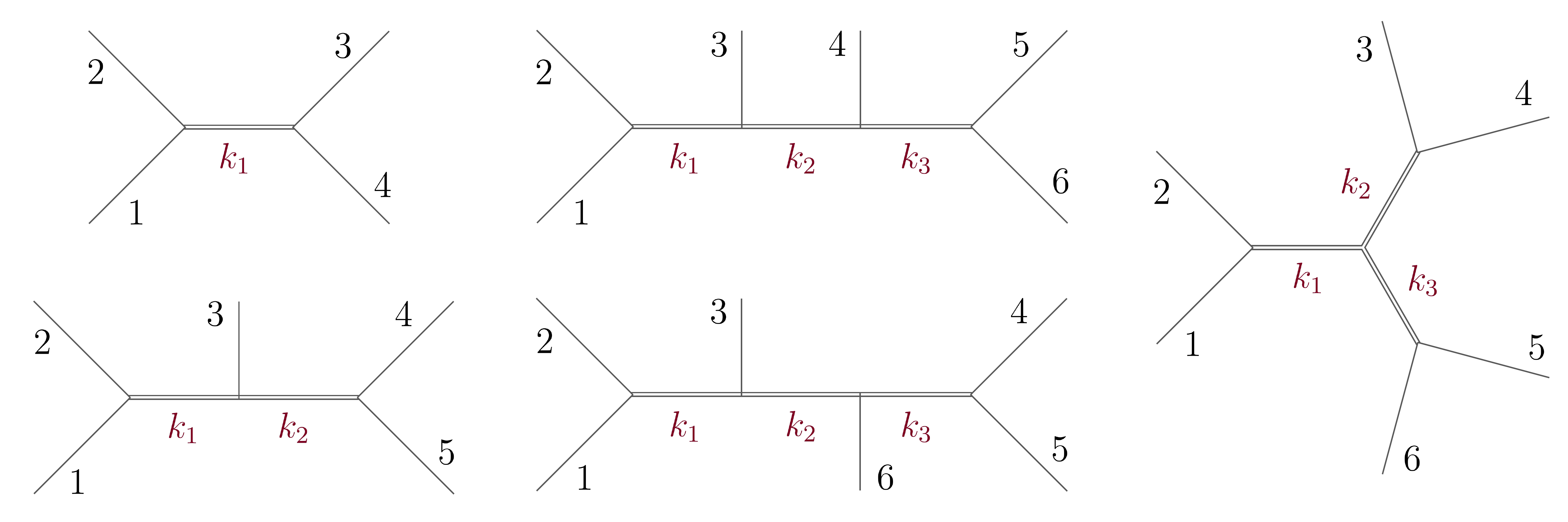}\vspace{-2mm}
	\caption{Consistent factorization in these channels is sufficient to rule out an enormous class of putative amplitudes. Counterclockwise from top left: four- and five-point half-ladder; six-point twisted half-ladder, star, and half-ladder.}
	\label{fig:factpic}
\end{figure}

In this paper we will argue that it is not, on account of the constraints imposed by factorization on {\it massive poles} onto a self-consistent set of three-point amplitudes.  This condition is amazingly restrictive.  Factorization in four-, five-, and six-point scattering is sufficient to vanquish the very simplest bespoke model as well as a large class of natural $P_n(u)$ constructions. These exclusions only require the consideration of scalars and vectors exchanged in the topologies depicted in \Fig{fig:factpic}.  Along the way, we will introduce useful techniques for computing residues on massive poles and summarize a new set of simple sum rules that directly constrain the residues and mass spectrum of any putative $n$-point amplitude.

Of course, the $n$-point amplitudes of string theory satisfy the constraints from factorization on massive poles.  But a corollary of this check is that we can directly extract the three-point amplitudes---coupling constants and all---directly from the higher-point amplitudes.

\medskip

\noindent {\bf Warmup.}---The constraining power of factorization is already evident at low point.  As a simple illustration of this approach,  let us briefly consider the tree-level tachyon amplitudes of the bosonic string.
These amplitudes famously exhibit a linear spectrum $m_k^2=k-1$, where $k$ is a nonnegative integer labeling each level of the string, which supports modes of spin $0\leq \ell \leq k$.  Factorizing the four-point string amplitude onto levels $k=0$ and $1$, we obtain the residues
\eq{
R_{0}=1 \;\;  {\rm and} \;\;
R_{1}=2 - X_{2,4} ,
}{R01}
where $X_{i,j}=(p_i + p_{i+1} +\cdots + p_{j-1})^2$ are the planar Mandelstam invariants in mostly-plus signature.
Here $R_0$ is generated by a scalar at level $k=0$ and $R_1$ is generated by a scalar and a vector at level $k=1$.

Next, we compare these residues against an {\it ansatz residue} derived from a set of ansatz three-point on-shell amplitudes $A^{\ell_1 \ell_2 \ell_3}_{k_1 k_2 k_3}$ describing the three-particle interaction of the states at levels $k_1, k_2, k_3$ of spins $\ell_1, \ell_2, \ell_3$.  Explicitly,
\be 
A^{000}_{000} \,{=}\,\lambda_{000}^{000},\;\;A_{100}^{000} \,{=}\,\lambda_{100}^{000},\;\;{\rm and}\;\;
A^{100}_{100} \,{=}\,\lambda_{100}^{100}(\epsilon_1 p_2).\label{eq:3pt1}
\ee 
We then extract the corresponding Feynman vertices and glue them together to obtain the ansatz residues, 
\eq{
\hspace{-1mm} R_{0}\,{=} \,(\lambda_{000}^{000})^2\;\, {\rm and}\;\,
R_{1}\,{=}\, (\lambda_{100}^{000})^2 {+} \frac{1}{4} (\lambda_{100}^{100})^2 (1{-}2 X_{2,4}) .\hspace{-1mm}
}{R01ansatz}
Equating Eqs.~\eqref{R01} and \eqref{R01ansatz}, we  thus determine---up to unphysical signs---the coupling constants of the theory. 

From squared coupling constants in \Eq{R01ansatz}, we see that the above construction is simply a restatement of the usual condition of positivity in four-point scattering.  However, since we have determined the three-point amplitudes, we can feed these results back into higher-point residues to obtain new bounds.

For example, the five-point residue of the string amplitude, at level $k\,{=}\,1$ for both internal propagators, is 
\eq{
R_{11} = 3 -X_{2,4} - X_{2,5} - X_{3,5} + X_{2,4} X_{3,5} ,
}{}
while the corresponding ansatz residue is
\eq{
\hspace{-3mm} R_{11} =&\phantom{{}{+}\,{}} (\lambda^{000}_{100})^2\lambda^{000}_{110} + \frac{1}{4} (\lambda^{100}_{100})^2 \lambda^{110}_{110} \\& {+}\,\frac{1}{4}\left[(\lambda^{100}_{100})^2 \lambda^{110}_{110} {-}2\lambda^{000}_{100}\lambda^{100}_{100}\lambda^{100}_{110}\right](X_{2,4}{+}X_{3,5})\\& {-}\,\frac{1}{2}(\lambda^{100}_{100})^2 \lambda^{110}_{110}X_{2,5} - \frac{1}{4}(\lambda^{100}_{100})^2 \lambda^{110\prime}_{110}X_{2,4}X_{3,5}.
}{R11ansatz}
Here $A^{000}_{100}$ and $A^{100}_{100}$ were fixed at four point, and we have defined the additional ansatz amplitudes,
\be 
\begin{aligned}
&A^{000}_{110} = \lambda^{000}_{110}, \quad
A^{100}_{110} = \lambda^{100}_{110}(\epsilon_1 p_2), \;\; {\rm and}\\
&A^{110}_{110} = \lambda^{110}_{110}(\epsilon_1 \epsilon_2) + \lambda^{110\prime}_{110} (\epsilon_1 p_2)(\epsilon_2 p_1).\label{eq:3pt2}
\end{aligned}
\ee 
The residues in \Eqs{R01ansatz}{R11ansatz} parameterize the consistent space allowed by factorization and our assumed spectrum.  Hence, they can be used to fix the couplings of the theory or, conversely, rule out a putative amplitude that is incompatible with factorization.  

\medskip

\noindent {\bf General Procedure.}---The above analysis generalizes to any spin and number of external particles.  As we will see, however,  factorization constraints from scalar and vector exchange in four-, five-, and six-point scattering are sufficient to dramatically reduce the space of consistent scattering amplitudes!  Let us now describe the general procedure for constructing these ansatz residues.

First, we take a chosen spectrum of masses and spins as an input.  For concreteness, we assume a spectrum that is continuously connected to that of the string: a single scalar at level $k=0$ and a single scalar and vector at level $k=1$, with no degeneracy.    However, we will assume nothing a priori about the precise values of the masses or the coupling constants of the theory. 
Throughout, we will consider amplitudes in which the external states are the colored scalars of mass $m_0^2$ at level $k=0$.

Second, we enumerate all three-point on-shell amplitudes for these states.
The on-shell three-point amplitudes for zero, one, or two external vectors were given in Eqs.~\eqref{eq:3pt1} and \eqref{eq:3pt2}.   
The amplitude for three vectors is \footnote{Note that the three-point amplitude $A^{\ell_1\ell_2\ell_3}$ has an intrinsic ordering in its external legs fixed by the color ordering of the diagram. Hence, the clockwise vs. counterclockwise ordering of legs on a given three-point amplitude is related by a possible sign flip.  In particular, a three-point amplitude picks up the  sign $(-1)^{\ell_1 +\ell_2 + \ell_3}$ when any two indices are swapped.}
\be 
\!\!\! A^{111}_{111} {=} \lambda^{111}_{111} [(\epsilon_1 \epsilon_2)\!(\epsilon_3 p_1) {+} {\rm cyc}]{+} \lambda^{111\prime}_{111}(\epsilon_1 p_2) \! (\epsilon_2 p_3) \!(\epsilon_3 p_1) .\!\!
\ee
Here the first and second terms correspond to the Yang-Mills cubic vertex and anomalous triple gauge boson vertex, respectively.
From the above amplitudes it is easy to derive the corresponding Feynman vertices.  

Finally, to build the ansatz residue we glue these Feynman vertices together with the Feynman propagators for scalars and vectors.  The numerator for the latter is $\Pi_{\mu\nu} = \eta_{\mu\nu} + p_\mu p_\nu/m_k^2$, where $p^2 = -m_k^2$ on the cut.
Note that there is no dependence on the spacetime dimension $D$ in the scalar and vector propagators.  Furthermore, traces of the metric do not appear at tree level, so the bounds we encounter will be $D$-independent.

\medskip

\noindent {\bf Examples.}---In this section we compute the residues for various higher-point amplitudes and compare them directly to the residues constructed using the general procedure outlined above.  As we will see, constraints from higher-point factorization are exceedingly powerful.

\medskip\noindent {\it String Amplitudes.}  To begin, consider the standard tree-level $n$-point Koba-Nielsen factor~\cite{GSW,Polchinski,BBS},
\be 
A_{\rm KN}(p_i) = \int \frac{d^{n} z}{{\rm SL}(2,\mathbb{R})} \frac{\prod_{i<j} (z_{j,i})^{-2p_i p_j}}{z_{1,2} \cdots z_{n,1}} 
, \label{eq:AV}
\ee
which we refer to as a ``string amplitude'' in a slight abuse of notation~\footnote{These worldsheet integrals play an important role in describing {\it all} open string amplitudes at tree level: all M{\"o}bius invariant integrals ``twisted'' by the Koba-Nielsen factor can be written as a linear combination of integrals of this form taken over a basis of $(n-3)!$ different color orderings. They are also directly related to famous amplitudes in string theory. For instance, with a simple kinematic shift, the all-tachyon amplitude of the bosonic string takes this form. In addition, this object has been proposed as a self-consistent theory in and of itself, dubbed $Z$-theory. Finally, at four points, the gluon amplitude in superstring theory is given by this factor times a simple prefactor $\mathcal{F}^4/u$.}.
Here $p_i$ are the external momenta, $z_i$ are moduli integrated over the disc, and $z_{i,j}= z_i-z_j$ are their differences.  The spectrum of states is defined by $m_k^2 = k$~\footnote{We work in units where the string scale is $\alpha^\prime = 1$}.  
Since the external legs of the string amplitude are planar-ordered, all factorization channels appear as simple poles in the planar Mandelstam invariants $X_{i,j}$ defined previously.  Conveniently, these $n(n-3)/2$ variables form a minimal basis for all $n$-point Mandelstam invariants.  To make this dependence manifest, we recast \Eq{eq:AV} in terms of the  SL(2,$\mathbb{R}$) invariant cross-ratios, $u_{i,j} = (z_{i-1,j} z_{i,j-1})/(z_{i,j}z_{i-1,j-1}) $, 
in which case the string amplitude becomes
\be 
A_{\rm KN}(X_{i,j}) = \int \frac{d^{n} z}{{\rm SL}(2,\mathbb{R})} \frac{\prod_{i<j} u_{i,j}^{X_{i,j}+\ao}}{z_{1,2}\cdots z_{n,1}} 
. \label{eq:Au}
\ee
Note that in going from \Eq{eq:AV} to \Eq{eq:Au} we have shifted the planar Mandelstam invariants by a constant Regge intercept $\ao$, so the amplitude automatically has singularities at $X_{i,j} = -m_k^2$ and the spectrum is $m_k^2 = k+\ao$.

The $u$ variables satisfy the nontrivial relations,
\eq{
u_{i,j} + \prod_{(i^\prime,j^\prime)  \cap (i,j)} u_{i^\prime,j^\prime} = 1,
}{u_eq}
where $(i',j')$ corresponds to all chords of the $n$-gon that intersect the chord $(i,j)$. These equations admit an $(n-3)$-dimensional space of solutions parameterized precisely by the moduli $z_i$ of the gauge-fixed worldsheet.

Using traditional $z_i$ variables,  it is challenging to extract the residues at arbitrary levels. But the raison d'\^{e}tre of the $u_{i,j}$ variables is to transparently manifest \textit{all} singularities, giving us a simple and explicit way to extract arbitrary residues. This is most easily done by a change of variables to \textit{positive coordinates}, where the amplitude takes the form 
\eq{
A_{\rm KN}(X_{i,j}) = \int_0^\infty \prod_{\mathcal{C}\in\mathcal{T}} \frac{dy_{\mathcal{C}}}{y_{\mathcal{C}}} \prod_{i<j} u_{i,j}^{X_{i,j}+\ao}.
}{string_can_form}
Here each $y_{\mathcal{C}}$ is associated with a chord $\mathcal{C}$ appearing in some choice of triangulation $\mathcal{T}$ of the $n$-gon. A simple algorithm expresses $u_{i,j}$ in terms of $y_{\mathcal{C}}$~\cite{Arkani-Hamed:2023lbd}.
In this form the singularities $X_{\mathcal{C}}+\alpha_0 \to 0$ of chords in the triangulation ${\cal T}$ are associated with the logarithmic divergence of the integral near $y_{\mathcal{C}} \to 0$. The integral diverges for $X_{\mathcal{C}} +\alpha_0<0$ and must be defined by analytic continuation, indeed developing poles when $X_{\mathcal{C}}+\alpha_0$ is a negative integer. For generic $X_{\mathcal{C}}+\alpha_0$ the integrand has branch cuts in $y_{\cal C}$~\footnote{Note that this integrand behavior is distinct from the integral, i.e., the amplitude itself, which is meromorphic in $X_{\cal C}$.}. However, when all $X_{\mathcal{C}} +\alpha_0 = -k_{\mathcal{C}}$, the integrand instead has poles at $y_{\mathcal{C}} \to 0$. The residue of the integral on the pole is computed by the residue of the integrand at $y_{\mathcal{C}} \to 0$.  

In summary, the algorithm for computing the residue associated with {\it any} diagram or triangulation ${\cal T}$ is then simple: use the positive parameterization associated with ${\cal T}$ in \Eq{string_can_form}, set the $X_{\mathcal{C}}+\alpha_0 \rightarrow -k_\mathcal{C}$, and compute the residue of the integrand at $y_{\mathcal{C}} \to 0$.

With this procedure we straightforwardly compute all residues of the four-, five-, and six-point string amplitudes on the maximal factorization channels on which all internal legs are localized to levels $k=0,1$.  We then compare these string residues to the ansatz residues constructed from gluing together arbitrary three-point amplitudes with real couplings.  Remarkably, for any value of the Regge intercept $\ao\geq-1$, the string amplitude is perfectly consistent with these factorization conditions.  As an immediate byproduct of this exercise, we extract the explicit three-point amplitudes of string theory for levels $k=0,1$ (i.e., masses $m^2 = \ao$ or $1+\ao$) and spins $\ell=0,1$, given in the table below. 
We constructed these couplings uniquely from consistent factorization of higher-point string amplitudes, without making use of worldsheet vertex operators.
It is straightforward to compute all three-point amplitudes of the string by generalizing this approach to higher levels, modulo considerations of degeneracy, which we leave to future work.

\begin{center}
\bgroup
\def\arraystretch{1.2}
\begin{tabular}{c | c || l}
$(k_1, k_2, k_3)$ & $(\ell_1,\ell_2,\ell_3)$ & three-point amplitude $A^{\ell_1 \ell_2 \ell_3}_{k_1 k_2 k_3}$ \\ \hline \hline
$(0,0,0)$ & $(0,0,0)$ &  $1$  \\ \hline
$(1,0,0)$ & $(0,0,0)$ &  $\sqrt{\dfrac{1+\ao}{2}}$ \\ \hline
$(1,0,0)$ & $(1,0,0)$ &  $\sqrt{2}\,(\epsilon_1  p_2 )$ \\ \hline
$(1,1,0)$ & $(0,0,0)$ & $1+\dfrac{\ao}{2}$\\ \hline
$(1,1,0)$ & $(1,0,0)$ & $\sqrt{1+\ao}\,(\epsilon_1  p_2)$ \\ \hline
$(1,1,0)$ & $(1,1,0)$ & $(\epsilon_1  \epsilon_2) - 2(\epsilon_1  p_2)(\epsilon_2  p_1)$\\ \hline
$(1,1,1)$ & $(0,0,0)$ & $\left(2+\dfrac{\ao}{2}\right)\sqrt{\dfrac{1+\ao}{2}}$\\ \hline
$(1,1,1)$ & $(1,0,0)$ & $\sqrt{2}\left(1+\dfrac{\ao}{2}\right) (\epsilon_1  p_2)$\\ \hline
$(1,1,1)$ & $(1,1,0)$ & $\sqrt{\dfrac{1+\ao}{2}}\left[(\epsilon_1  \epsilon_2)\,{-}\, 2(\epsilon_1  p_3)(\epsilon_2  p_3)\right]$\\ \hline
$(1,1,1)$ & $(1,1,1)$ & $\begin{aligned}\sqrt{2}\left[(\epsilon_1  \epsilon_2)(\epsilon_3  p_1) + {\rm cyclic}\right]  \\+ 2 \sqrt{2} (\epsilon_1  p_2)(\epsilon_2  p_3)(\epsilon_3  p_1)\end{aligned}$
\end{tabular}
\egroup
\end{center}

\medskip\noindent {\it Bespoke Amplitudes.} Next, let us demonstrate how consistent factorization imposes stringent constraints on the bespoke  amplitudes defined in Ref.~\cite{Cheung:2023uwn}.  For concreteness, we focus on the very simplest version of these amplitudes with a nonlinear spectrum, corresponding to $h=2$ in the nomenclature of that paper. 

For this amplitude, the spectrum of resonances at level $k$ is defined by the nonlinear rational polynomial $m_k^2 = (k^2\,{+}\,p_1 k\,{+}\,p_2)/(k\,{+}\,q_2)$,
where $p_1,p_2,q_2$ are parameters satisfying $p_1 q_2\,{-}\,p_2\,{>}\,0$.
The corresponding $n$-point bespoke amplitude is simply a sum over the usual string amplitudes,\eq{
A(X_I) = 2^{-n(n-3)/2} J^n \sum_{\alpha_I \in \{\pm\}} A_{\rm KN}({-}\nu_{\alpha_I}({-}X_I)),
}{A_bespoke}
 where $I\,{=}\,1,{\ldots},n(n{-}3)/2$ and in each term the planar Mandelstam $X_I$ has been composed with a nonlinear function
$\nu_\pm (\mu) =  [\mu - p_1 \pm \sqrt{(\mu-p_1)^2 + 4(q_2 \mu-p_2)}]/2$.
The average over square root branches, for each index $\alpha_1\ldots,\alpha_{n(n-3)/2}$, ensures that the resulting amplitude is free of branch cuts and has polynomial residues. 
For appropriate choices of $p_1,p_2,q_2$, Ref.~\cite{Cheung:2023uwn} showed that the four-point amplitude possesses a dual resonant representation and is consistent with partial wave unitarity.
The factor $J =  \sqrt{h^{-1} (dm^2_k/dk)}|_{k=0} =  \sqrt{(p_1 q_2{-}p_2)/2q_2^2}$ fixes the canonical normalization of level $k=0$ external states, so the bespoke amplitude automatically factorizes on the exchange of those states.  Proper factorization on states at higher $k$, however, is not yet guaranteed.

The residues of the bespoke amplitudes are computed by extracting the residues of the string amplitude using \Eq{string_can_form} and inserting them into \Eq{A_bespoke}.   Comparing the resulting bespoke residues against the ansatz residues at four, five, and six point for level $k=0,1$, we find that modulo some trivial theories with purely scalar interactions, there are {\it no values} of $p_1, p_2, q_2$ that are consistent with factorization. While a one-parameter family with $p_1/2 =2 q_2 = 1+p_2/q_2$  does indeed satisfy the factorization constraints for the half-ladders, it  fails for the six-point twisted half-ladder depicted in \Fig{fig:factpic}.

Thus, the very simplest bespoke amplitudes are inconsistent with factorization onto a single scalar at level $k=0$ and a single scalar and vector at $k=1$.  
It remains an open question whether these bounds can be evaded in the broader class of bespoke amplitudes \cite{Cheung:2023uwn}, or perhaps by including degeneracy.

\medskip\noindent {\it Deformed String Integrand.}  The critical feature of \Eq{string_can_form} is that it makes factorization on the level $k=0$ poles obvious. When $X_{i^\star,j^\star} + \alpha_0\to 0$ the integral develops a pole near $u_{i^\star,j^\star} \to 0$. Because of the ``binary'' character of Eq.~\eqref{u_eq}, the limit $u_{i^\star,j^\star} \to 0$ sends $u_{i^\prime,j^\prime} \to 1$ for the chords $(i^\prime,j^\prime)$ that cross $(i^\star,j^\star)$.  Hence $\prod_{i,j} u_{i,j}^{X_{i,j} + \alpha_0}$ factorizes in a way that mirrors the amplitude. 
This suggests a more general class of functions that exhibit level $k=0$ factorization:  simply multiply the $n$-point integrand by a form factor $P_n(u)$ engineered to factorize into the product of lower-point factors $P_{n_L}(u_L)P_{n_R}(u_R)$ when $u_{X}= 0$. It is natural to restrict to a finite polynomial $P(u)$ to preserve the Regge behavior characteristic of string amplitudes. 

Let us define a simple family of deformations, parameterized by an integer $m\geq3$ that defines an $m$-gon $Q^{(m)}$ inside the momentum $n$-gon. We then define
\be
P^{(m)}_n (u) = \prod_{Q^{(m)}} \bigg[1 + g \!\! \prod_{(i,j) \in Q^{(m)}} \! (1-u_{i,j})\bigg],
\ee
where $(i,j)$ sums over all pairs of vertices in $Q^{(m)}$ and we take $u_{i,i+1}=0$. The parameter $g$ controls the deformation away from the original string amplitude.

Note that if a chord $X$ intersects $Q^{(m)}$, it must intersect at least one of the chords inside $Q^{(m)}$. Therefore, on a factorization channel where $u_X \to 0$, the product of $(1 - u_{i,j})$ vanishes on account of \Eq{u_eq}, leaving us with the product over all the $Q^{(m)}$ that do not intersect $X$, so factorization at level $k=0$ then follows.

For our analysis, we consider the $P^{(3)}_n(u)$ and $P^{(4)}_n(u)$ amplitudes, which we dub the ``triangle'' and ``quadrilateral'' deformations, respectively. In both cases, three-point amplitudes at level $k=0,1$ are completely fixed by four- and five-point scattering.  Feeding these couplings back into higher-point amplitudes, we find that if $g\neq0$ then the six-point residues for both the triangle and quadrilateral are consistent for the \mbox{four-,} five-, and six-point half-ladders but {\it inconsistent} for the twisted half-ladder topology shown in \Fig{fig:factpic}.  Hence, both the triangle and quadrilateral models are ruled out~\footnote{Consistent factorization technically allows for a  triangle deformation with the peculiar choice of $g=-1$.  For this choice, all four-, five-, and six-point residues vanish at levels $k=0,1$, which is as far as we have taken our analysis.  Nevertheless, this setup is ruled out by four-point partial wave unitarity at level $k=3$.}.

Amusingly, the triangle deformation can actually be ruled out at four point via partial wave unitarity, though at great cost.
Computing the partial wave expansion on the Gegenbauer polynomials, we find that for nonzero $g$, the first negative partial wave is at spin $0$ or $1$, but at very high level $k\sim 1/g$.  This dramatic difference in effort illustrates the power of higher-point factorization.

\medskip

\noindent {\bf Sum Rules.}---Higher-point factorization imposes a set of nonlinear sum rules that depend on the masses and the numerical coefficients of the residues of a putative amplitude.  These sum rules have the distinct advantage that they can be used to cleanly eliminate inconsistent theories without resorting to the laborious procedure of gluing together ansatz three-point amplitudes.

To this end, let us define a {\it general parameterization} of the residues output by the ansatz residue for arbitrary couplings.
For example, the general four-point half-ladder residues at level $k=0,1$ are
\eq{
R_{0}  = \omega_{0}^{0}  \;\; {\rm and} \;\;
R_{1}  = \omega_{1}^{0} + \omega_{1}^{1}X_{2,4} ,
}{4ptRes}
while at five point they are
\eq{
R_{00}   =& \phantom{ {}+{} } \omega_{00}^{000} \\
R_{10}   =& \phantom{ {}+{} } \omega_{10}^{000}+ \omega_{10}^{100}X_{2,4}  \\
R_{01}   =& \phantom{ {}+{} } \omega_{01}^{000} + \omega_{01}^{001}X_{3,5}  \\
R_{11}   =& \phantom{ {}+{} } \omega_{11}^{000}+ \omega_{11}^{100}X_{2,4} + \omega
_{11}^{010}X_{2,5} \\&  + \omega_{11}^{001}X_{3,5}  + \omega
_{11}^{101}X_{2,4} X_{3,5} .
}{5ptRes}
Here $n$-point residues are fully characterized by the numerical coefficients $\omega_{k_1 k_2  {\cdots} }^{r_1 r_2 \cdots}$, where the subscripts $k_1, k_2, \ldots$ denote the levels on which each internal leg has been localized and the superscripts $r_1, r_2, \ldots$ denote the power of each Mandelstam invariant $X_I$.  The ordering of $I$ is defined by the lexicographic order of the $i,j$ subscripts in $ X_{i,j} = X_{I}$ that appear in a given residue~\footnote{For example, for the five-point half-ladder diagram, we are factorizing on the $(X_{1,3},X_{1,4})=(-m_{k_1}^2,-m_{k_2}^2)$ channel. The residues can depend on the remaining planar Mandelstam variables, which in lexicographic order are $(X_{2,4},X_{2,5},X_{3,5})$. On the $k_1=k_2=1$ channel, the coefficient of $X_{2,4}X_{3,5}$ is $\omega_{11}^{101}$. In App.~\ref{app:6pt}, we give the six-point sum rules for all topologies. For example, for the six-point half-ladder, we factorize on the $(X_{1,3}, X_{1,4},X_{1,5})=(-m_{k_1}^2,-m_{k_2}^2,-m_{k_3}^2)$ channel. The remaining planar Mandelstam variables, in lexicographic order, are $(X_{2,4},X_{2,5},X_{2,6},X_{3,5},X_{3,6},X_{4,6})$, so the coefficient, in the $(k_1,k_2,k_3)=(1,1,0)$ residue, of $X_{2,4}X_{3,5}$ is $\omega_{110}^{100100}$.  Note that the sum of the numbers of lower and upper indices always equals the total number of planar Mandelstam variables, $n(n-3)/2$.}.    Note that the $\omega$ coefficients satisfy  identities required by the flip isometries, so for example, $\omega_{k_1 k_2}^{r_1 r_2 r_3} = \omega_{k_2 k_1}^{r_3 r_2 r_1}$.
 
Since the ansatz residue is constructed by gluing together arbitrary three-point amplitudes, it is a polynomial in $X_I$ with very precise relations linking the numerical coefficients of each kinematic structure.  These conditions correspond to a set of nonlinear sum rules that relate the $\omega$ coefficients with the mass spectrum $m^2_k$, which we now summarize for four-, five-, and six-point residues computed at levels $k=0,1$.

\medskip
\noindent {\it Four-Point Sum Rules.} For the general four-point residue in \Eq{4ptRes}, we can compare to the ansatz residue to extract the corresponding three-point amplitudes, 
\eq{
A_{000}^{000} &= (\omega_0^{0})^{ 1/2} \\
A_{100}^{000} &= [\omega_1^{0}-(4 m^2_0 -m^2_1 )\omega_1^{1}/2]^{1/2} \\
A_{100}^{100} &= (-2\omega_1^{1})^{1/2} (\epsilon_1 p_2) .
}{A3_omega1}
Immediately, the reality of the coupling constants imply inequalities on the residue coefficients, 
\be 
\omega_0^{0}\,{>}\,0, \quad \omega_1^{1}\,{<}\,0,\quad \textrm{and} \quad \omega_1^{0}\,{\geq}\, \frac{4 m^2_0 - m^2_1}{2}\omega_1^{1},\label{sumrule1}
\ee
where the first and second inequalities are not saturated because we assume that there are nonzero couplings to both the scalar and the vector.

\medskip
\noindent {\it Five-Point Sum Rules.} To obtain the sum rules at five point, we plug the three-point amplitudes in \Eq{A3_omega1} back into the ansatz residue and again compare with the general residue in \Eq{5ptRes}.  Doing so fixes several more three-point amplitudes,
\be 
\hspace{-2mm} 
\begin{aligned}
A_{110}^{000} &\,{=}\,\frac{4 \omega_{11}^{000} \! {-}12 m^2_0 \omega_{11}^{001} \!{+}(2 m^2_1{-}9m_0^2) \omega_{11}^{010}\! {+}9 m^4_0 \omega_{11}^{101} }{2(2\omega_{1}^{0}-4 m^2_0\omega_{1}^{1}+m^2_1\omega_{1}^{1})}\\
A_{110}^{100} &\,{=}\, \frac{-2 \omega_{11}^{001}- \omega_{11}^{010} +3 m^2_0 \omega_{11}^{101} }{(-\omega_1^{1})^{1/2}( 2\omega_{1}^{0}-4 m^2_0\omega_{1}^{1}+m^2_1\omega_{1}^{1})^{1/2}} (\epsilon_1 p_2)\\
A_{110}^{110} &\,{=}\, \frac{\omega_{11}^{010}}{\omega_1^{1}} (\epsilon_1 \epsilon_2)+ \frac{2\omega_{11}^{101}}{\omega_1^{1}}(\epsilon_1 p_2)(\epsilon_2 p_1) ,\label{A3_omega2}
\end{aligned}\hspace{-3mm}
\ee
as well as the sum rules,
\be 
(\omega_{00}^{000},\;\omega_{01}^{000},\;\omega_{01}^{001} )= (\omega_{0}^{0})^{1/2}\times (\omega_{0}^{0},\; \omega_{1}^{0},\; \omega_{1}^{1}).
\label{sumrule2}
\ee
See App.~\ref{app:6pt} for the analogous sum rules for consistent factorization at six point.

\medskip

\noindent {\bf Discussion.}---In this paper we have argued that consistent factorization is a remarkably stringent constraint on multiparticle scattering.  Our analysis is by no means exhaustive, but rather intended as an invitation to explore this rich set of novel higher-point consistency conditions.   There are many questions directly spurred by our preliminary observations.

For example, it is well motivated to ask whether there are any known deformations of the $n$-point amplitude that actually satisfy consistent factorization.  Some natural candidates include bespoke amplitudes with more highly nonlinear spectra, as well as other $P(u)$ deformations of the string integrand.

A convenient byproduct of our analysis is that we can mechanically extract the three-point couplings of the string directly from the $n$-point scattering amplitudes.  It would be very interesting to use this tool to learn about the structure of interactions and density of states of the string directly from the amplitudes themselves.   These and related lines of investigation hold the promise of fundamental new insights into the question of what---if anything---makes string theory special.

\vspace{5mm}

\noindent {\it Acknowledgments:} 
We thank David Gross and Aaron Hillman for useful discussions.
N.A.H. is supported by the DOE (Grant No.~DE-SC0009988), by the Simons Collaboration on Celestial Holography, and further support was made possible
by the Carl B. Feinberg cross-disciplinary program in innovation at the IAS. C.C. is supported by the DOE (Grant No.~DE-SC0011632) and by the Walter Burke Institute for Theoretical Physics. C.F. is supported by FCT/Portugal (Grant No.~2023.01221.BD).
G.N.R. is supported by the James Arthur Postdoctoral Fellowship at New York University.

\vspace{80mm}

\onecolumngrid

\section{Six-Point Sum Rules}\label{app:6pt}

We construct general sum rules for consistent factorization of all six-point diagrams at levels $k=0,1$. For ease of use, we also provide these results in a supplementary text file.

We first perform the factorization calculation for the six-point half-ladder diagram depicted in Fig.~2 of the Letter by computing the glued residues and comparing against an arbitrary six-point ansatz residue parameterized by the $\omega$ coefficients,
\be 
\begin{aligned}
R_{000}   =& \phantom{ {}+{} } \omega_{000}^{000000} \\
R_{100} =& \phantom{ {}+{} } \omega_{100}^{000000}+\omega_{100}^{100000}X_{2,4} \\
R_{010} =& \phantom{ {}+{} } \omega_{010}^{000000}+\omega_{010}^{000100}X_{3,5} \\
R_{001} =& \phantom{ {}+{} } \omega_{001}^{000000}+\omega_{001}^{000001}X_{4,6} \\
R_{110} =& \phantom{ {}+{} } \omega_{110}^{000000}+\omega_{110}^{100000}X_{2,4}+\omega_{110}^{010000}X_{2,5}+ \omega_{110}^{000100}X_{3,5} + \omega_{110}^{100100}X_{2,4}X_{3,5} \\
R_{101} =& \phantom{ {}+{} } \omega_{101}^{000000}+\omega_{101}^{100000}X_{2,4}+\omega_{101}^{000001}X_{4,6} + \omega_{101}^{100001}X_{2,4}X_{4,6} \\
R_{011} =& \phantom{ {}+{} } \omega_{011}^{000000}+\omega_{011}^{000100}X_{3,5}+\omega_{011}^{000010}X_{3,6}+ \omega_{011}^{000001}X_{4,6} + \omega_{011}^{000101}X_{3,5}X_{4,6} \\
R_{111} =& \phantom{ {}+{} } \omega_{111}^{000000}+\omega_{111}^{100000}X_{2,4}+\omega_{111}^{010000}X_{2,5}+\omega_{111}^{001000}X_{2,6}+
\omega_{111}^{000100}X_{3,5}+\omega_{111}^{000010}X_{3,6}+\omega_{111}^{000001}X_{4,6} \\&+ \omega_{111}^{100100}X_{2,4}X_{3,5} + \omega_{111}^{100010}X_{2,4}X_{3,6} + \omega_{111}^{100001}X_{2,4}X_{4,6}+ \omega_{111}^{010001}X_{2,5}X_{4,6}+ \omega_{111}^{000101}X_{3,5}X_{4,6}\\&+ \omega_{111}^{100101}X_{2,4}X_{3,5}X_{4,6},
\end{aligned}
\ee
which satisfy the flip isometries $\omega_{k_1 k_2 k_3}^{r_1 r_2 r_3 r_4 r_5 r_6} = \omega_{k_3 k_2 k_1}^{r_6 r_5 r_3 r_4 r_2 r_1}$.
Here, we are factorizing on the $(X_{1,3},X_{1,4},X_{1,5})=(-m_{k_1}^2, -m_{k_2}^2, -m_{k_3}^2)$ channel.
We find mass-independent sum rules,
\be 
\begin{aligned} 
\omega_{000}^{000000} &=(\omega_{0}^0)^2 \qquad &   \omega_{011}^{000000}&=(\omega_{0}^0)^{1/2} \omega_{11}^{000} \qquad &\omega_1^1\omega_{111}^{001000} &=(\omega_{11}^{010})^2 \\
 \omega_{001}^{000000} &=\omega_{0}^0 \omega_{1}^0 &    \omega_{011}^{000100}&=(\omega_{0}^0)^{1/2} \omega_{11}^{001} &  \omega_1^1 \omega_{111}^{000100} &=(\omega_{11}^{001})^2\\
   \omega_{001}^{000001}&=\omega_{0}^0 \omega_{1}^1     &\omega_{011}^{000010}&=(\omega_{0}^0)^{1/2} \omega_{11}^{010}&\omega_1^1\omega_{111}^{000010} &= \omega_{11}^{010}\omega_{11}^{001} \\
    \omega_{010}^{000000}&=\omega_{0}^0 \omega_{1}^0 &  \omega_{011}^{000001}&=(\omega_{0}^0)^{1/2} \omega_{11}^{001}& \omega_1^1\omega_{111}^{010001} &=\omega_{11}^{010}\omega_{11}^{101}\\
    \omega_{010}^{000100}&=\omega_{0}^0 \omega_{1}^1 &   \omega_{011}^{000101}&=(\omega_{0}^0)^{1/2} \omega_{11}^{101} &  \omega_1^1 \omega_{111}^{000101} &=\omega_{11}^{001}\omega_{11}^{101}   \\
  \omega_{101}^{000000} &=(\omega_{1}^0)^2    &  && \omega_1^1\omega_{111}^{100101} &= (\omega_{11}^{101})^2, \\
    \omega_{101}^{000001}&=\omega_{1}^0 \omega_{1}^1 \\
    \omega_{101}^{100001}&=(\omega_{1}^1)^2 
\end{aligned}
\ee
along with mass-dependent sum rules,
\be
\begin{aligned}
0=&\phantom{{}\,{+}\;{}}\omega_1^0 \left[(m_1^2-m_0^2)^2(\omega_{11}^{010})^2 -2m_0^2(m_0^2+m_1^2)\omega_{11}^{010}\omega_{11}^{001} + m_0^2(m_0^2-4m_1^2)(\omega_{11}^{001})^2\right] \\& - 2\omega_1^1 \left[m_1^2 (\omega_{11}^{000})^2 +m_0^6 (\omega_{11}^{010})^2 -m_0^2(2m_0^4-3m_0^2 m_1^2+m_1^4)\omega_{11}^{010}\omega_{11}^{001}\right.
\\& \qquad\qquad \left. +m_0^2(m_1^2-m_0^2)^2(\omega_{11}^{001})^2 + m_1^2(m_1^2-3m_0^2) \omega_{11}^{000}\omega_{11}^{010} -3 m_0^2 m_1^2 \omega_{11}^{000}\omega_{11}^{001}\right] \\& +m_1^2\omega_1^1 [2\omega_1^0+(m_1^2-4m_0^2)\omega_1^1]\omega_{111}^{000000}
\\ \\
0=&\phantom{{}\,{+}\;{}}\omega_1^0 \left[(m_0^2+m_0^2)(\omega_{11}^{010})^2 +m_0^2 (m_0^2-4m_1^2)\omega_{11}^{001}\omega_{11}^{101} + (2m_1^2-m_0^2)\omega_{11}^{010}\omega_{11}^{001}-m_0^2(m_0^2+m_1^2)\omega_{11}^{010}\omega_{11}^{101}\right] \\& - \omega_1^1 \left[2m_0^2(\omega_{11}^{010})^2+m_1^2\omega_{11}^{000}\omega_{11}^{010}+2m_1^2\omega_{11}^{000}\omega_{11}^{001}-3m_0^2 m_1^2 \omega_{11}^{000}\omega_{11}^{101}-3m_0^2 m_1^2 (\omega_{11}^{001})^2\right.
\\& \qquad\qquad \left. -2m_0^4\omega_{11}^{010}\omega_{11}^{001}+ 2m_0^2(m_1^2-m_0^2)^2 \omega_{11}^{001}\omega_{11}^{101}+m_0^2(m_1^2-m_0^2)(2m_0^2-m_1^2) \omega_{11}^{010}\omega_{11}^{101} \right] \\& + m_1^2\omega_1^1 [2\omega_1^0+(m_1^2-4m_0^2)\omega_1^1]\omega_{111}^{000001} \\ \\
0=&\phantom{{}\,{+}\;{}}\omega_1^0 \left[(\omega_{11}^{010})^2 +2(2m_1^2-m_0^2)\omega_{11}^{010}\omega_{11}^{101} + m_0^2 (m_0^2-4m_1^2)(\omega_{11}^{101})^2 \right] \\& - 2\omega_1^1 \left[m_0^2(\omega_{11}^{010})^2+m_1^2(\omega_{11}^{001})^2-3m_0^2 m_1^2\omega_{11}^{001}\omega_{11}^{101}+m_0^2(m_1^2-m_0^2)^2(\omega_{11}^{101})^2 \right.
\\& \qquad\qquad \left. +m_1^2\omega_{111}^{100}\omega_{11}^{010}+(m_1^2-m_0^2)(2m_0^2-m_1^2)\omega_{11}^{010}\omega_{11}^{101} \right] \\& + m_1^2\omega_1^1 [2\omega_1^0+(m_1^2-4m_0^2)\omega_1^1]\omega_{111}^{100001}.
\end{aligned} 
\ee

\pagebreak

For the six-point twisted half-ladder diagram shown in Fig.~2, we define a new ansatz residue $\widetilde R$ in terms of coefficients $\widetilde\omega_{k_1 k_2 k_3}^{r_1 r_2 r_3 r_4 r_5 r_6}$,
\be 
\begin{aligned}
\widetilde R_{000}   =& \phantom{ {}+{} } \widetilde\omega_{000}^{000000} \\
\widetilde R_{100} =& \phantom{ {}+{} } \widetilde \omega_{100}^{000000}+\widetilde \omega_{100}^{010000}X_{2,4} \\
\widetilde R_{010} =& \phantom{ {}+{} } \widetilde \omega_{010}^{000000}+\widetilde \omega_{010}^{000001}X_{3,6} \\
\widetilde R_{001} =& \phantom{ {}+{} } \widetilde \omega_{001}^{000000}+\widetilde \omega_{001}^{100000}X_{1,5} \\
\widetilde R_{110} =& \phantom{ {}+{} } \widetilde \omega_{110}^{000000}+\widetilde \omega_{110}^{010000}X_{2,4}+\widetilde \omega_{110}^{000100}X_{2,6}+ \widetilde\omega_{110}^{000001}X_{3,6} + \widetilde\omega_{110}^{010001}X_{2,4}X_{3,6} \\
\widetilde R_{101} =& \phantom{ {}+{} } \widetilde\omega_{101}^{000000}+\widetilde \omega_{101}^{100000}X_{1,5}+\widetilde \omega_{101}^{010000}X_{2,4}+\widetilde\omega_{101}^{110000}X_{1,5}X_{2,4} \\
\widetilde R_{011} =& \phantom{ {}+{} } \widetilde \omega_{011}^{000000}+\widetilde \omega_{011}^{100000}X_{1,5}+\widetilde \omega_{011}^{000010}X_{3,5}+ \widetilde \omega_{011}^{000001}X_{3,6} + \widetilde \omega_{011}^{100001}X_{1,5}X_{3,6} \\
\widetilde R_{111} =& \phantom{ {}+{} } \widetilde \omega_{111}^{000000}+\widetilde \omega_{111}^{100000}X_{1,5}+\widetilde \omega_{111}^{010000}X_{2,4}+\widetilde \omega_{111}^{001000}X_{2,5}+
\widetilde \omega_{111}^{000100}X_{2,6}+\widetilde\omega_{111}^{000010}X_{3,5}+\widetilde \omega_{111}^{000001}X_{3,6} \\&+ \widetilde \omega_{111}^{110000}X_{1,5}X_{2,4} + \widetilde \omega_{111}^{100100}X_{1,5}X_{2,6} + \widetilde\omega_{111}^{100001}X_{1,5}X_{3,6}+ \widetilde\omega_{111}^{010010}X_{2,4}X_{3,5}+ \widetilde \omega_{111}^{010001}X_{2,4}X_{3,6}\\&+ \widetilde\omega_{111}^{110001}X_{1,5}X_{2,4}X_{3,6},
\end{aligned}
\ee
which satisfy the rotation isometries $\widetilde\omega_{k_1 k_2 k_3}^{r_1 r_2 r_3 r_4 r_5 r_6} = \widetilde\omega_{k_3 k_2 k_1}^{r_2 r_1 r_3 r_5 r_4 r_6}$. 
Here, we are factorizing on the $(X_{1,3},X_{1,4},X_{4,6})=(-m_{k_1}^2,-m_{k_2}^2,-m_{k_3}^2)$ channel.
We find mass-independent sum rules relating the half-ladder and twisted half-ladder coefficients, 
\be
\begin{aligned}
\widetilde\omega_{000}^{000000}&= (\omega_0^0)^2 \qquad &\widetilde\omega_{011}^{000000} &= (\omega_0^0)^{1/2} \omega_{11}^{000} \qquad & \omega_1^1 \widetilde\omega_{111}^{001000}  &= (\omega_{11}^{010})^2 \\
\widetilde\omega_{001}^{000000} &= \omega_0^0 \omega_1^0& \widetilde\omega_{011}^{100000} &= (\omega_0^0)^{1/2} \omega_{11}^{001}  \qquad & \omega_1^1 \widetilde \omega_{111}^{000100}&=\omega_{11}^{010}\omega_{11}^{001} \\
\widetilde\omega_{001}^{100000} &=\omega_0^0 \omega_1^1  & \widetilde\omega_{011}^{000010} &= (\omega_0^0)^{1/2} \omega_{11}^{010} \qquad & \omega_1^1 \widetilde\omega_{111}^{000001}&=(\omega_{11}^{001})^2 \\
\widetilde\omega_{010}^{000000} &= \omega_0^0 \omega_1^0& \widetilde\omega_{011}^{000001} &= (\omega_0^0)^{1/2} \omega_{11}^{001}  \qquad & \omega_1^1 \widetilde\omega_{111}^{100100} &=\omega_{11}^{010}\omega_{11}^{101} \\
\widetilde\omega_{010}^{000001} &= \omega_0^0 \omega_1^1 & \widetilde\omega_{011}^{100001} &=  (\omega_0^0)^{1/2} \omega_{11}^{101} \qquad & \omega_1^1\widetilde\omega_{111}^{100001}&=\omega_{11}^{001}\omega_{11}^{101} \\
\widetilde\omega_{101}^{000000} &= (\omega_1^0)^2 && \qquad & \omega_1^1 \omega_{111}^{110001}&=(\omega_{11}^{101})^2, \\
\widetilde\omega_{101}^{100000} &= \omega_1^0 \omega_1^1 \\
\widetilde\omega_{101}^{110000} &= (\omega_1^1)^2
\end{aligned} 
\ee
along with mass-dependent sum rules,
\be
\begin{aligned}
0=&\phantom{{}\,{+}\;{}}\omega_1^0 \left[(m_0^4+2m_0^2m_1^2-m_1^4)(\omega_{11}^{010})^2 +2m_0^2(3m_1^2-m_0^2)\omega_{11}^{010}\omega_{11}^{001} + (m_0^4+2m_1^2)(\omega_{11}^{001})^2\right] \\& + \omega_1^1 \left[2m_1^2 (\omega_{11}^{000})^2 +m_0^4(m_1^2-2m_0^2)(\omega_{11}^{010})^2 -4m_0^4(m_1^2-m_0^2)\omega_{11}^{010}\omega_{11}^{001}\right.
\\& \qquad\qquad \left. +(m_1^2 -m_0^2)^2(m_1^2-2m_0^2)(\omega_{11}^{001})^2 +2 m_1^2(m_1^2-3m_0^2) \omega_{11}^{000}\omega_{11}^{010} -6 m_0^2 m_1^2 \omega_{11}^{000}\omega_{11}^{001}\right] \\& -m_1^2\omega_1^1 [2\omega_1^0+(m_1^2-4m_0^2)\omega_1^1]\widetilde\omega_{111}^{000000}
\\ \\
0=&\phantom{{}\,{+}\;{}}\omega_1^0 \left[(m_1^2-m_0^2)(\omega_{11}^{010})^2 -(m_0^4+2m_1^4)\omega_{11}^{001}\omega_{11}^{101} + m_0^2\omega_{11}^{010}\omega_{11}^{001}+m_0^2(m_0^2-3m_1^2)\omega_{11}^{010}\omega_{11}^{101}\right] \\& + \omega_1^1 \left[m_0^2(2m_0^2-m_1^2)(\omega_{11}^{010})^2-m_1^2\omega_{11}^{000}\omega_{11}^{010}-2m_1^2\omega_{11}^{000}\omega_{11}^{001}+3m_0^2 m_1^2 \omega_{11}^{000}\omega_{11}^{101}+3m_0^2 m_1^2 (\omega_{11}^{001})^2\right.
\\& \qquad\qquad \left.+(m_1^2-m_0^2)^2(2m_0^2-m_1^2) \omega_{11}^{001}\omega_{11}^{101} - (2m_0^4-5m_0^2 m_1^2+m_1^4)\omega_{11}^{010}\omega_{11}^{001} +2m_0^4(m_1^2-m_0^2)\omega_{11}^{010}\omega_{11}^{101}\right] \\& +m_1^2\omega_1^1 [2\omega_1^0+(m_1^2-4m_0^2)\omega_1^1]\widetilde\omega_{111}^{100000} \\ \\
0=&\phantom{{}\,{+}\;{}}\omega_1^0 \left[(\omega_{11}^{010})^2 -2m_0^2\omega_{11}^{010}\omega_{11}^{101} + (m_0^4+2m_1^4)(\omega_{11}^{101})^2 \right] \\& + \omega_1^1 \left[(m_1^2-2m_0^2)(\omega_{11}^{010})^2+2m_1^2(\omega_{11}^{001})^2-6m_0^2 m_1^2\omega_{11}^{001}\omega_{11}^{101}+(m_1^2-m_0^2)(m_1^2-2m_0^2)(\omega_{11}^{101})^2 \right.
\\& \qquad\qquad \left. +2m_1^2\omega_{111}^{100}\omega_{11}^{010}-4m_0^2(m_1^2-m_0^2)\omega_{11}^{010}\omega_{11}^{101} \right] \\& -m_1^2\omega_1^1 [2\omega_1^0+(m_1^2-4m_0^2)\omega_1^1]\widetilde\omega_{111}^{110000}.
\end{aligned} 
\ee

\pagebreak

For the final remaining topology, the six-point star diagram in Fig.~2, we define an ansatz  residue $\Rs$ in terms of coefficients $\os_{k_1 k_2 k_3}^{r_1 r_2 r_3 r_4 r_5 r_6}$,
\be 
\begin{aligned}
\Rs_{000}   =& \phantom{ {}+{} } \os_{000}^{000000} \\
\Rs_{100} =& \phantom{ {}+{} } \os_{100}^{000000}+\os_{100}^{001000}X_{2,5} \\
\Rs_{010} =& \phantom{ {}+{} } \os \omega_{010}^{000000}+\os_{010}^{100000}X_{1,4} \\
\Rs_{001} =& \phantom{ {}+{} } \os_{001}^{000000}+\os_{001}^{000010}X_{3,6} \\
\Rs_{110} =& \phantom{ {}+{} } \os_{110}^{000000}+\os_{110}^{100000}X_{1,4}+\os_{110}^{010000}X_{2,4}+ \os_{110}^{001000}X_{2,5} + \os_{110}^{101000}X_{1,4}X_{2,5} \\
\Rs_{101} =& \phantom{ {}+{} } \os_{101}^{000000}+\os_{101}^{001000}X_{2,5}+\os_{101}^{000100}X_{2,6}+\os_{101}^{000010}X_{3,6} + \os_{101}^{001010}X_{2,5}X_{3,6} \\
\Rs_{011} =& \phantom{ {}+{} } \os_{011}^{000000}+\os_{011}^{100000}X_{1,4}+\os_{011}^{000010}X_{3,6}+ \os_{011}^{000001}X_{4,6} + \os_{011}^{100010}X_{1,4}X_{3,6} \\
\Rs_{111} =& \phantom{ {}+{} } \os_{111}^{000000}+\os_{111}^{100000}X_{1,4}+\os_{111}^{010000}X_{2,4}+\os_{111}^{001000}X_{2,5}+\os_{111}^{000100}X_{2,6}+\os_{111}^{000010}X_{3,6}+\os_{111}^{000001}X_{4,6} \\&+\os_{111}^{101000}X_{1,4}X_{2,5} + \os_{111}^{100100}X_{1,4}X_{2,6}+\os_{111}^{100010}X_{1,4}X_{3,6} + \os_{111}^{010010}X_{2,4}X_{3,6} + \os_{111}^{001010}X_{2,5}X_{3,6}\\& +\os_{111}^{001001}X_{2,5}X_{4,6} + \os_{111}^{101010}X_{1,4}X_{2,5}X_{3,6},
\end{aligned}
\ee
which satisfy the rotation and flip isometries $\os_{k_1 k_2 k_3}^{r_1 r_2 r_3 r_4 r_5 r_6} = \os_{k_3 k_1 k_2}^{r_3 r_4 r_5 r_6 r_1 r_2}=\os_{k_3 k_2 k_1}^{r_1 r_6 r_5 r_4 r_3 r_2}$.
Here, we are factorizing on the $(X_{1,3},X_{3,5},X_{1,5})=(-m_{k_1}^2,-m_{k_2}^2,-m_{k_3}^2)$ channel.
Due to the enhanced symmetries of the star diagram, consistent factorization imposes a shorter list of sum rules,
\be
\begin{aligned}
\os_{000}^{000000} &= (\omega_0^0)^2 \qquad & \os_{011}^{000000} &= (\omega_0^0)^{1/2} \omega_{11}^{000} \\
\os_{001}^{000000} &= \omega_0^0 \omega_1^0 & \os_{011}^{000001} &= (\omega_0^0)^{1/2} \omega_{11}^{010} \\
\os_{001}^{000010} &= \omega_0^0 \omega_1^1 & \os_{011}^{100000} &= (\omega_0^0)^{1/2}\omega_{11}^{001} \\
 & & \os_{011}^{100010} &=(\omega_0^0)^{1/2} \omega_{11}^{101}.
\end{aligned} 
\ee

\twocolumngrid

\bibliographystyle{utphys-modified}
\bibliography{gluing}

\end{document}